\documentclass[conference]{IEEEtran}
\IEEEoverridecommandlockouts
\usepackage{cite}
\usepackage{amsmath,amssymb,amsfonts}
\usepackage{algorithmicx}
\usepackage{graphicx}
\usepackage{textcomp}
\usepackage{xcolor}
\usepackage{caption}
\usepackage{graphicx}
\usepackage{array}
\usepackage{booktabs}
\usepackage{caption}
\usepackage{multirow}
\usepackage{algorithm}
\usepackage{algpseudocode}
\usepackage{amsmath}
\usepackage[a4paper, 
    top=0.75in,    
    bottom=1.73in,  
    left=0.65in,   
    right=0.7in,  
    footskip=0.8in 
]{geometry}

\renewcommand{\arraystretch}{1.5} 
\usepackage{tabularx, colortbl, xcolor}
\def\BibTeX{{\rm B\kern-.05em{\sc i\kern-.025em b}\kern-.08em
    T\kern-.1667em\lower.7ex\hbox{E}\kern-.125emX}}
\begin{document}

\title{Building Network Digital Twins Part II: Real-Time Adaptive PID for Enhanced State Synchronization \\
}

\author{\IEEEauthorblockN{1$^{st}$ John Sengendo}
\IEEEauthorblockA{Dept. of Information Engineering and Computer Science\\
University of Trento\\
Email: john.sengendo@unitn.it}
\and
\IEEEauthorblockN{2$^{nd}$ Fabrizio Granelli}
\IEEEauthorblockA{Dept. of Information Engineering and Computer Science\\
University of Trento\\
Email: fabrizio.granelli@unitn.it}
}

\maketitle

\begin{abstract}
As we evolve towards more heterogeneous and cutting-edge mobile networks, Network Digital Twins (NDTs) are proving to be a promising paradigm in solving challenges faced by network operators, as they give a possibility of replicating the physical network operations and testing scenarios separately without interfering with the live network. However, with mobile networks becoming increasingly dynamic and heterogeneous due to massive device connectivity, replicating traffic and having NDTs synchronized in real-time with the physical network remains a challenge, thus necessitating the need to develop real-time adaptive mechanisms to bridge this gap. In this part II of our work, we implement a novel framework that integrates an adaptive Proportional–Integral–Derivative (PID) controller to dynamically improve synchronization. Additionally, through an interactive user interface, results of our enhanced approach demonstrate an improvement in real-time traffic synchronization.
\end{abstract}

\section{Introduction}
In addition to the transition toward sophisticated and large-scale networks such as 6G, efficient management of mobile networks remains a key challenge for network operators, due to factors such as network heterogeneity, underlying hardware complexity, virtualization technologies, and AI-driven services. Besides advanced techniques such as Software-Defined Networking (SDN) and Network Function Virtualization (NFV), which are applied in 5G networks to enable services such as network slicing \cite{b1}, recent technology that has emerged is the development of digital replicas known as Digital Twins (DTs), in the context of networks that are termed as Network Digital Twins (NDTs), replicating physical network systems \cite{b2}. These models facilitate remote real-time monitoring, predictive analysis and closed-loop automation \cite{b2}, with the ability to make decisions in real-time or near real-time thus increasing resiliency of networks. Additionally, a future state traffic predictive capability discussed in \cite{b3} is another important use case for NDTs that enables seamless network management. 

NDTs further aid in mitigating congestion, optimizing routing, and enhancing quality of service (QoS) by precisely estimating traffic patterns within the network. The coupling of the physical twin (PT) and digital twin (DT) or in other words ensuring that both entities remain in ”sync” is significant because any error in alignment can cause traffic interruption within the network especially when the DT controls the PT. This pivotal area of state synchronization or alignment thus forms the main aim of our study in this paper.

\begin{figure}[h]
    \centering
    \includegraphics[width=0.5\textwidth, height=0.31\textheight]{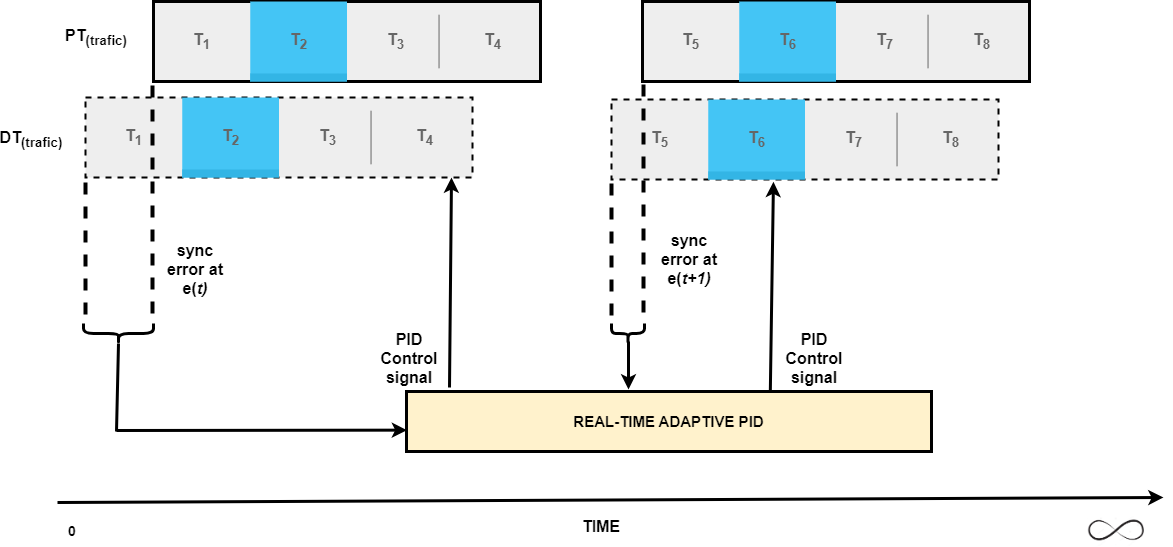} 
    \caption{Synchronization illustration of a Digital Twin with a Physical Twin using a real-tme adaptive PID}
    \label{fig:example} 
\end{figure}

In the preliminary study of this work (Part I) \cite{b4}, we gave an introduction on solving synchronization with a combined approach of a Long short-term memory (LSTM) neural network with a manual PID to reduce prediction error. Given the dynamic nature of networks, in part II of our work we enhance our approach by implementing a real-time adaptive PID technique, with the capability of dynamically tuning in real-time to enhance synchronization between the digital and physical twins network traffic. Our study in this paper is further motivated by the highly cognitive nature of cutting-edge networks like 6G, which demand real-time adaptive and responsive mechanisms \cite{b5}.
In Figure 1, we give an illustration of the time series nature of network traffic demonstrating its continuous evolvement over time and how misalignment with the DT can exist. This variability requires an adaptive mechanism to continuously minimize ”sync” errors $E(t)$, $E(t+1)$ between the PT and DT at different time steps.

Furthermore, since performance speed and lower inference time are key in time series scenarios like network traffic, in this study we applied a Convolutional neural network (CNN) based approach, which offers faster and parallel processioning \cite{b6} in comparison to LSTM approach as highlighted in Table I. Additionally, to further provide a real-time possibility for monitoring traffic alignment in real-time, we implemented a user interface for visualization and on-the-fly tuning of the PID parameters, offering a comprehensive framework for deeper insights into the synchronization process.
\renewcommand{\arraystretch}{1.0} 
\begin{table}[h]
    \centering
    \scriptsize 
    \begin{tabular}{|c|c|p{1.6cm}|p{1.5cm}|p{1.4cm}|}
        \hline
        Part & Model & Processing Type & Speed Performance & Inference Time \\
        \hline
        I & LSTM & Sequential & Slower & Higher \\
        II & CNN & Parallel & Faster & Lower \\
        \hline
    \end{tabular}
    \caption{Comparison of LSTM and CNN \cite{b6} \cite{b7}}
    \label{tab:part1_part2_comparison}
\end{table}

The rest of the paper is organized as follows: Section II
presents the related works on NDTs, highlighting the current progress on traffic prediction for state synchronization. Section III describes the system
model, which includes the experimental setup and implementation framework. Section IV gives a summary of the results obtained. Section V concludes this paper.

\section{RELATED WORKS}

Traffic prediction is fundamental for NDTs, enabling intelligent network management and resource allocation. Recent research efforts have explored a number of techniques such as artificial intelligence (AI)-driven approaches to enhance traffic prediction and state synchronization in digital twin environments, which are crucial in ensuring that NDTs accurately reflect the physical network conditions. In this section, we provide related research which has been published relevant to the subject.

In \cite{b2}, authors present a survey on DTs for networks, providing a state-of-the-art about the technology. In their literature, they highlight on how synchronization forms the pivot for a seamless operation between the PT and DT. Along the same lines, in \cite{b3} authors propose an efficient data-driven framework that adapts to varying network tasks and traffic types, demonstrating an improvement in prediction accuracy by integrating multiple learning models tailored for diverse network conditions. In more recent studies, many have underscored the growing role of AI and machine learning (ML) techniques in enhancing operations and synchronization accuracy. For example,  work in \cite{b8} emphasizes on the role of ML and deep learning in enhancing the efficiency of Digital Twin Networks (DTN). In their work they elaborate on how this integration can improve accuracy of predictive models running in DT systems which further can enhance the twins alignment. 

Furthermore, research provided in \cite{b9} employs deep Q-Learning techniques and generative adversarial networks (GANs) to enhance traffic prediction in dynamic network environments. Their approach provides a framework for addressing fluctuating traffic flows by continuously refining prediction models based on real-time feedback.
Exploring more literature, in \cite{b10} authors explore game-theoretic approaches for vehicular digital twin synchronization. In the approach they propose, the model applied optimizes network selection and minimizes latency through cooperative game-based strategies, ensuring real-time data consistency in connected vehicle environments. Moreover, authors in \cite{b11} formulate the DT synchronization problem analytically,  by further demonstrating that state-dependent synchronization policies provide near-optimal performance. Their framework works on optimizing update frequency to counter-balance accuracy and resource efficiency.
To further investigate synchronization in dual digital twins, the study presented in \cite{b12} proposes a Lyapunov stability-based synchronization method to advance real-time synchronization. The main focus in their framework is in improving modeling accuracy in energy applications such as EV battery management.

While these studies underscore contributions towards the advancement of synchronization, none have fully resolved the inherent challenge of achieving real-time synchronization especially with the highly dynamic, and sometimes rapidly changing conditions in mobile networks which demand adaptive mechanisms, thus remains an open problem to address. In the next sections of our work, we provide experimental implementations of our approach, further demonstrating how the adaptive PID-based technique significantly improves synchronization.

\begin{figure}[h]
    \centering
    \includegraphics[width=0.5\textwidth]{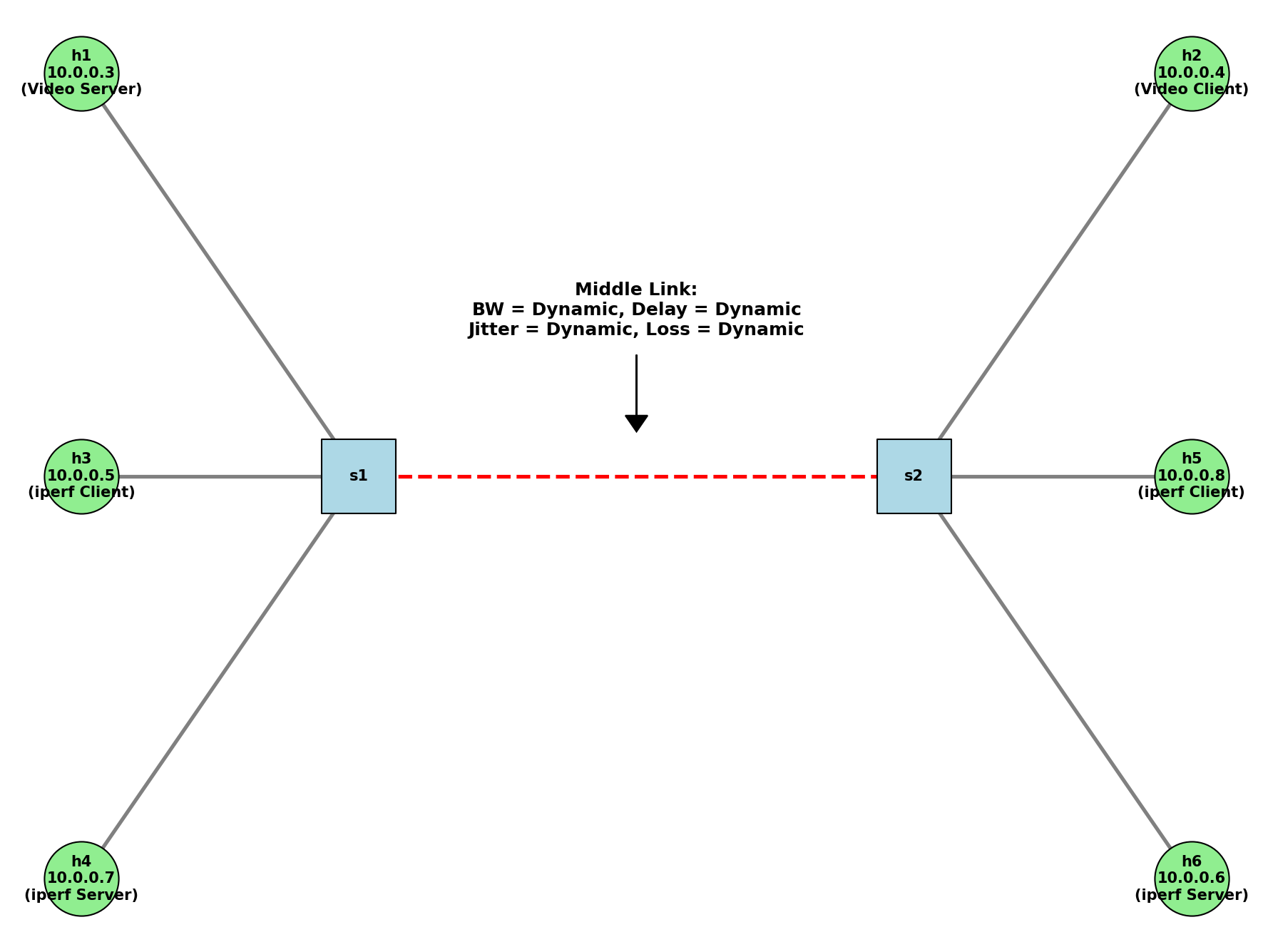} 
    \caption{Network topology}
    \label{fig:example} 
\end{figure}

\begin{figure*}[ht]
\vspace{0.1in} 
\centering
\includegraphics[width=0.99\textwidth, height=0.3\textheight]{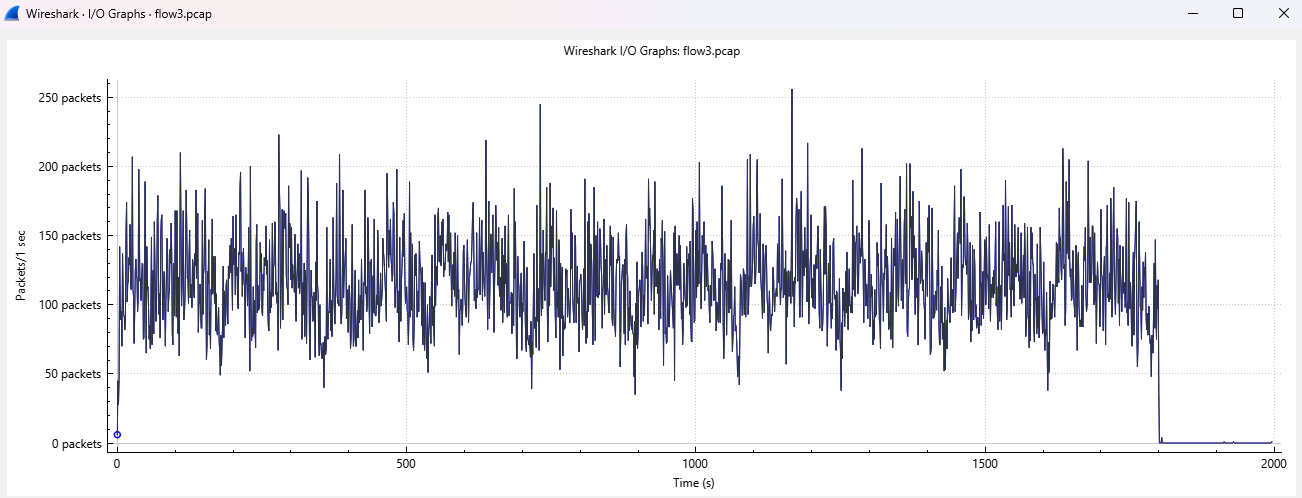}
\caption{Video streaming traffic}
\label{fig}
\end{figure*}

\begin{figure*}[ht]
\vspace{0.1in} 
\centering
\includegraphics[width=0.99\textwidth, height=0.3\textheight]{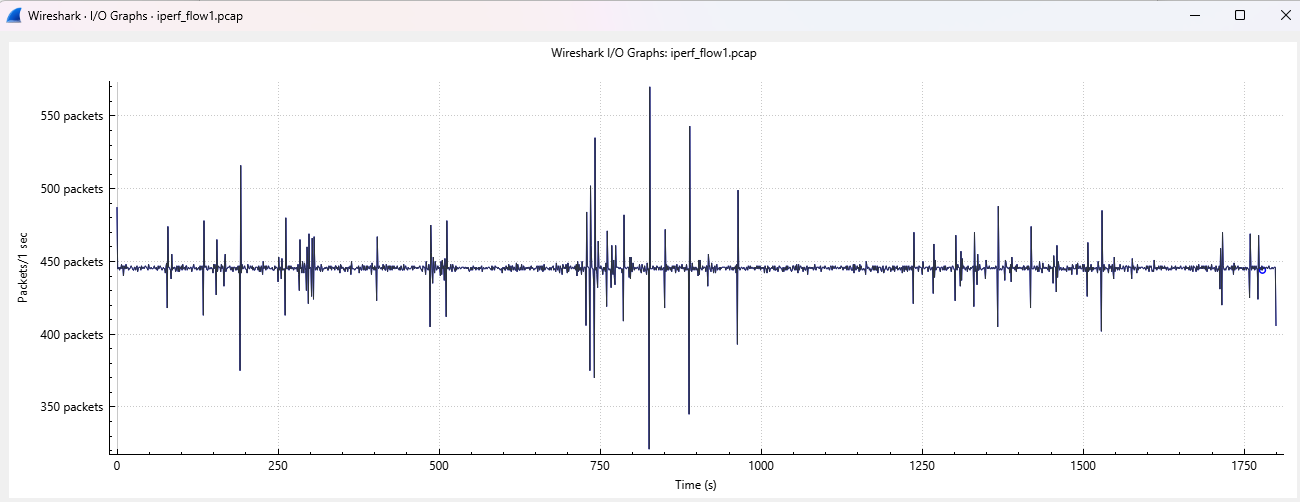}
\caption{iPerf traffic flow 1}
\label{fig}
\end{figure*}

\begin{table*}[t]
    \centering
    \renewcommand{\arraystretch}{0.95} 
    \resizebox{\textwidth}{!}{%
    \begin{tabular}{|l|c|c||l|c|}
        \hline
        \multicolumn{3}{|c||}{\textbf{Data Splitting and Training Settings}} & \multicolumn{2}{c|}{\textbf{CNN Model Architecture and Metrics}} \\
        \hline
        \textbf{Parameter} & \textbf{Value / Formula} & \textbf{Portion (\%)} & \textbf{Component} & \textbf{Details} \\
        \hline
        Data Split (Training) & First 45\% of data & 45\% & Model Type & CNN \\
        \hline
        Data Split (Validation) & Next 27.5\% of data & 27.5\% & Input Channels & 1 \\
        \hline
        Data Split (Test) & Last 27.5\% of data & 27.5\% & Conv Layers & 3 layers, kernel size 4, 'same' padding \\
        \hline
        BATCH\_SIZE & 32 & - & Batch Normalization & After each conv layer \\
        \hline
        EPOCHS & 50 & - & Error Metrics & MAE, RMSE \\
        \hline
        LEARNING\_RATE & 0.0001 & - &  &  \\
        \hline
    \end{tabular}%
    }
    \caption{Overview of the CNN model architecture, training settings, and data partitioning}
    \label{tab:cnn_overview}
\end{table*}

\section{Experimental Implementation}

This section presents the experimental setup and methodology used, elaborating more on the experimentation design to capture and predict traffic while integrating an adaptive PID controller to enhance synchronization.

\subsection{Experimental Setup}
Our experiment was carried out in ComNetsEmu \cite{b13} a network emulation tool that extends Mininet to support container-based network emulations using Docker. It gives capability of experimenting with Software-Defined Networking (SDN), Network Function Virtualization (NFV) in a flexible and scalable environment allowing for testing real network applications with greater realism \cite{b13}. In Figure 2, we present the topology setup where we adopted a dumbbell network topology consisting two switches s1 and s2 connecting six nodes categorized as follows:

\begin{itemize}
    \item \textbf{Video streaming nodes}: Depicted at the upper part of the topology were used for streaming a video service for 30 minutes, one acting as a server (h1) and the other as a client (h2) thus generating real-time video traffic.
    \item \textbf{iPerf servers}: Depicted at the middle and lower parts of the topology, running iPerf to generate additional network traffic flows, h4 being a server to client (h5) and h6 being a server to client (h3). In our experiment, both the video and iPerf traffic were running simultaneously for 30 minutes emulating a realistic multi-flow and bidirectional network traffic scenario.
\end{itemize}

\subsection{Data Collection}
To analyze the network traffic, transmitted data for the different flows was collected following procedures below:
\begin{itemize}
    \item The video streaming and iPerf traffic were captured at the bottleneck (middle) link using tcpdump, and the resulting pcap files were analyzed. Figures 3 illustrates the video traffic flow and Figure 4 one of the iPerf flows Flow (1) between h3 and h6. The analysis in the results section primarily focuses on the video flow due to its higher temporal variability and complex burst patterns, which offer more significant insights into network behavior under dynamic traffic conditions. The iPerf flow is presented for comparative context, as it generally demonstrates stable throughput characteristics typical of bulk data transfers.
    \item The video stream flow was pre-processed by converting the pcap file into transmitted packets per second in csv format, necessary for model training.
\end{itemize}
\subsection{Traffic Prediction model}
As mentioned earlier, a Convolutional Neural Network (CNN) model was used in our framework due to their superior efficiency in time series tasks, offering faster processing and lower inference time while maintaining comparable accuracy \cite{b14}. This is achieved by leveraging parallel convolutional operations, in contrast to the sequential processing required by recurrent neural networks such as LSTM and Gated recurrent units (GRUs) \cite{b14}. The selected training parameters and data splitting strategy as shown in Table II were well-suited for time series forecasting. The first and largest sequential portion of the data was used for training, ensuring the model learns temporal patterns effectively. The remaining data was partitioned into validation and testing sets in chronological order, supporting robust model tuning and performance evaluation on future, unseen data. The chosen batch size, number of epochs, and learning rate enabled stable convergence and efficient learning throughout the training process, especially in leveraging GPU acceleration for faster computation.

A window-based approach was applied where historical data (packet per second) was used as input features.  The prediction framework was to predict future traffic for a given time-steps into the future /time-steps ahead.
    To assess the effectiveness during training, Mean Absolute Error (MAE) and Root Mean Squared Error (RMSE) performance metrics were used, MAE measuring the average size of prediction errors, indicating overall accuracy without considering error direction \cite{b15} and RMSE taking account of larger errors by squaring deviations before averaging and taking the square root \cite{b15}.

\subsection{Adaptive PID integration for synchronization enhancement}
Following \textbf{Algorithm 1}, the adaptive PID was integrated into the framework. When the PID was activated with a click on the orange button on the user interface in Figure 6, the PID controller dynamically adjusts the predicted traffic in real-time based on traffic deviations.
\begin{algorithm}[H]
\caption{Real-time Network Digital Twin Traffic Synchronization with Adaptive PID Control}
\label{alg:dt-pid}
\begin{algorithmic}[1]  
    \Require Real-time traffic data stream
    \Ensure Traffic flow prediction with adaptive PID control optimization

    \State \textbf{Initialization:} Load pre-trained traffic prediction model \( M \)
    \State Initialize PID parameters \( K_p, K_i, K_d \)
    \State Set initial PID mode to \texttt{Deactivated}

    \State \textbf{Real-time prediction and control loop:}
    \While {system is operational}
        \State Capture and pre-process traffic data over window \( T_{\text{window}} \)
        \State Predict traffic flow:  
        \[
        \hat{T}(t) = M(T_{\text{window}})
        \]
        \State Compute prediction error:  
        \[
        e(t) = T_{\text{actual}}(t) - \hat{T}(t)
        \]

        \State Check if PID Activation is enabled 
        \If {PID mode is activated}
            \State Apply adaptive PID correction:  
            \[
            u(t) = K_p e(t) + K_i \int e(t) dt + K_d \frac{d}{dt} e(t)
            \]
            \State Apply control signal \( u(t) \) to adjust synchronization
        \EndIf

        \State Update digital twin and user interface in real time
    \EndWhile

    \State \textbf{Termination:} Stop execution upon user intervention
    
\end{algorithmic}
\end{algorithm}

\subsection{Error $e(t)$ correction and PID adaptation mechanism}
At every time instant $t$, the deviation between actual and predicted traffic is computed. This error/deviation  is fed into the PID controller following Equation 1, to generate a PID control signal $u(t)$. The key parameters \( K_p \),\( K_i \) and \( K_d \)  that contribute to $u(t)$ are elaborated below:
\begin{equation}
    u(t) = K_p e(t) + K_i \int e(t) \, dt + K_d \frac{de(t)}{dt}
\end{equation}

\begin{itemize}
    \item \( K_p \): The proportional gain determines the response of the controller to the current error. A higher \( K_p \) increases the correction based on the immediate value of the error \( e(t) \) \cite{b16}.

    \item \( K_i \): The integral gain addresses the accumulation of past errors by integrating \( e(t) \) over time helping to reduce residual errors that may persist due to steady-state deviations \cite{b16}.

    \item \( K_d \): The derivative gain anticipates future errors by considering the rate of change of \( e(t) \) helping to stabilize the system by reducing oscillations and providing damping \cite{b16}.
    \item Real-time tuning of these PID parameters which we made possible in our framework as shown on the bottom left of Figure 6, ensures optimal real-time performance.

\end{itemize}
\begin{figure*}\setcounter{figure}{5}
    \centering
    \includegraphics[width=1\textwidth, height=7.9cm]{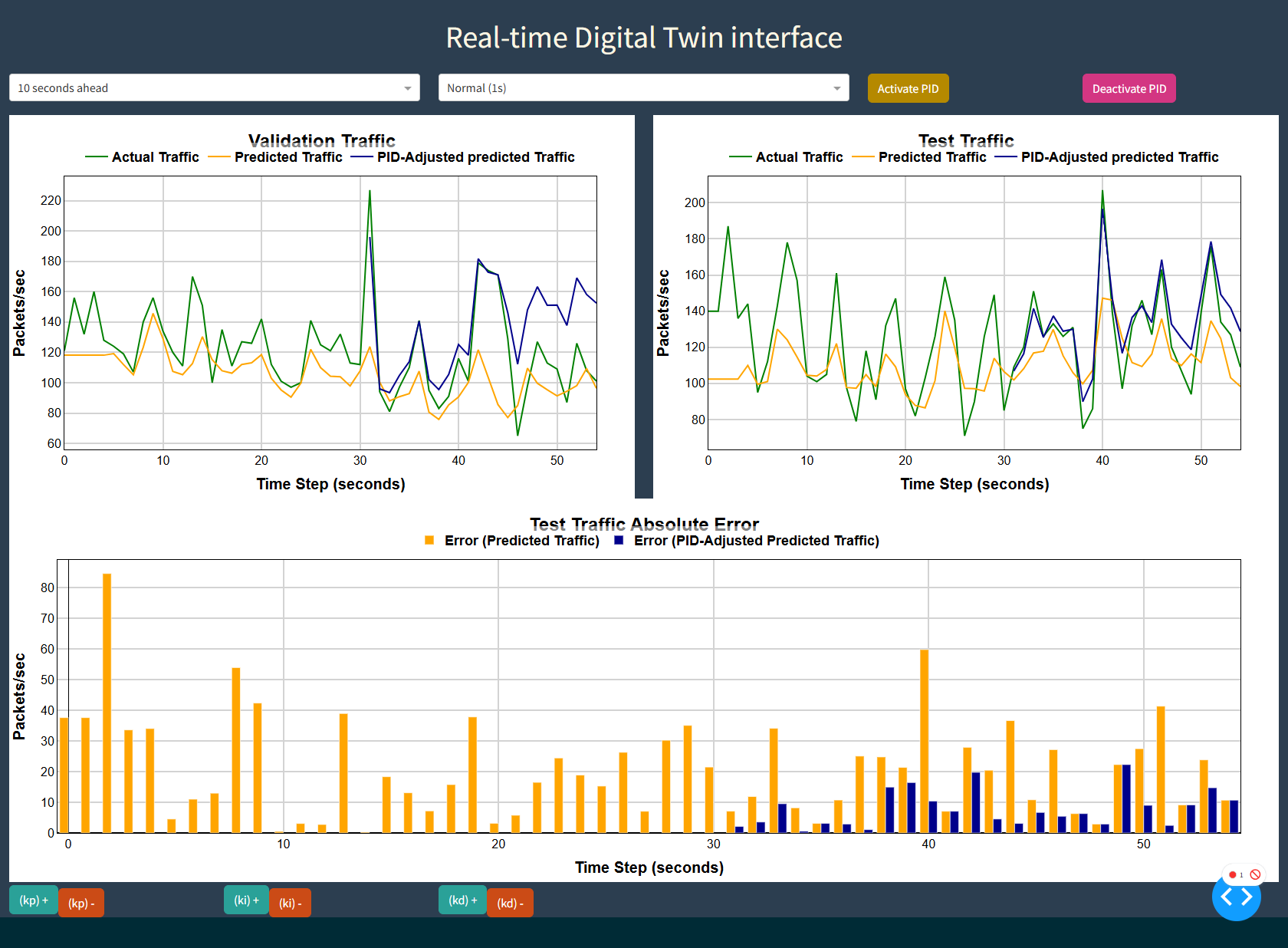} 
    \caption{Live Digital Twin user interface with real-time visualizations of traffic}
    \label{fig:example}
\end{figure*}

\section{Results Evaluation}

As mentioned earlier, the effectiveness of our DT framework was evaluated using the video streaming traffic. Dash, a Python framework for interactive web applications \cite{b17} was applied into our framework to develop a real-time user interface. The evaluation focused on prediction accuracy, synchronization, and responsiveness under dynamic traffic conditions. As illustrated in Figure 6, our developed live DT user interface enables real-time visualization of both validation and test traffic, alongside their respective predicted and PID-adjusted trajectories. From this interface, Figures 5 and 7 were extracted to visualize the test traffic and associated absolute errors respectively as they changed in real-time.

\subsection{Improved traffic synchronization}

In Figure 5, the predicted traffic without PID correction is depicted in orange, the actual traffic in green and the PID adjusted predictions in blue. As the traffic evolved overtime, we activated the PID just after 30 seconds. The PID adjusted predicted traffic maintained a close temporal alignment with the actual traffic in green, unlike the baseline predictions in orange which frequently diverges. This closer alignment is attributed to the feedback loop introduced by the PID controller, which ensures real-time adaptation to evolving traffic dynamics.

\subsection{Prediction error reduction}

To assess the error in synchronization as traffic evolved in real-time, the absolute prediction error was compared between the baseline prediction error and the PID-adjusted variant. As shown in Figure 7, following activation of the PID controller after $t = 30$ seconds, the PID-adjusted predictions consistently exhibited a lower error as the corresponding error bars in blue indicate a marked reduction in prediction error relative to the baseline predictions (orange bars), confirming the adaptive PID controller’s ability to suppress error and improve synchronization in real-time.

\setcounter{figure}{4} 
\begin{figure}[H]
\centering
\includegraphics[width=\columnwidth, height=4.5cm]{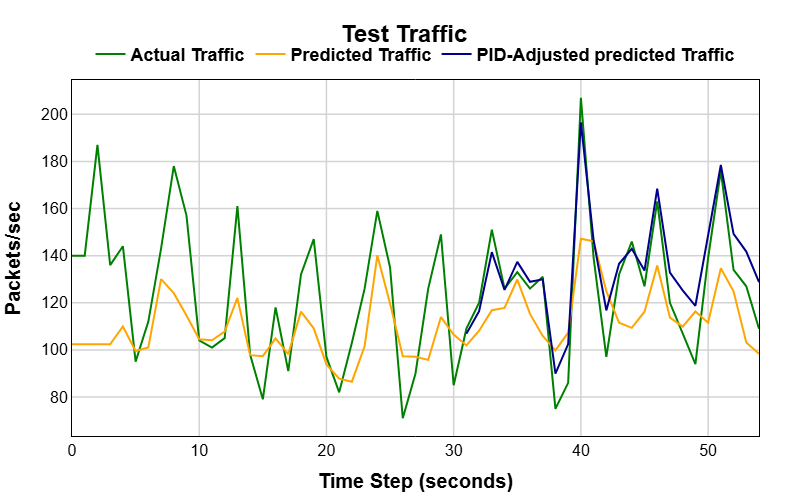}
\caption{Test traffic visualization depicting improved performance with PID adjusted predicted traffic closely synchronizing with actual traffic}
\label{fig:pid_traffic}
\end{figure}

\setcounter{figure}{6}
\begin{figure*}
\centering
\includegraphics[width=1\textwidth, height=6.2cm]{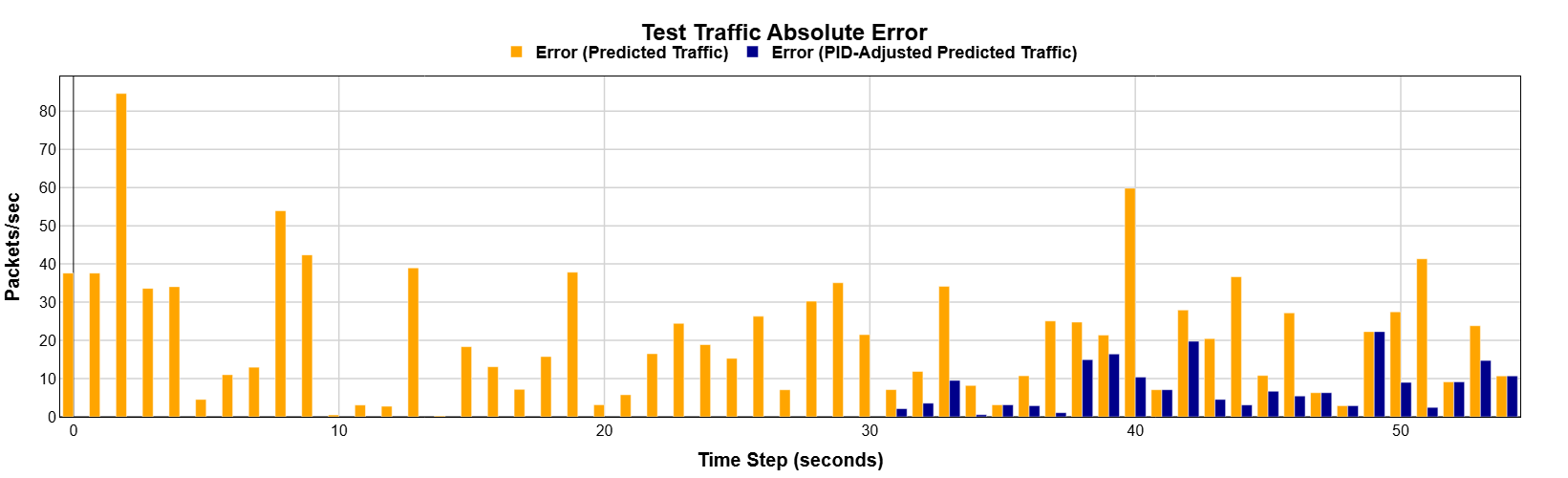}
\caption{Real-time error analysis depicting improved performance with lower error for PID adjusted predictions}
\label{fig}
\end{figure*}
\section{Conclusion}
Our work demonstrated the effectiveness an adaptive PID controller in enhancing network traffic prediction by reducing errors thus improving synchronization. The integration of real-time visualization, interactive controls, and fine-tuning capabilities provided a robust framework for traffic monitoring and adaptive adjustments. Results showed that our approach significantly improves synchronization in real-time.
Further research which we plan to present in part III will explore a more advanced approach integrating the capability of the Digital Twin to control the Physical Twin enabling orchestration a key feature in forthcoming 6G networks.

\section*{Acknowledgment}

This research received partial funding from the HORSE (Holistic, omnipresent, resilient services for future 6G wireless and computing ecosystems) project, supported by the Smart Networks and Services Joint Undertaking (SNS JU).

\vspace{12pt}
\color{red}

\end{document}